\documentclass[fleqn,twoside]{article}
\usepackage{espcrc2}
\usepackage{amssymb}
\usepackage{graphicx}
\newcommand{\ThickBase}{205\ }
\newcommand{\ThickEmu}{44\ }
\newcommand{\spacebeforefigurecaption}{\vspace{-6.5mm}}
\newcommand{\spaceafterfigurecaption}{\vspace{-8.5mm}}
\title{Fast automated scanning of OPERA emulsion films}

\author{
         G. Sirri \address{Istituto Nazionale di Fisica Nucleare, viale Berti Pichat 6/2, 40127 Bologna, Italy}
         oh behalf of the OPERA Collaboration.
}

\begin{document}
%______________________________________________________________________________
%

\begin{abstract}
\begin{minipage}[]{17cm}
{\hspace{-4.5mm} \footnotesize Presented at the 10$^\textrm{th}$ Topical Seminar on Innovative Particle and Radiation Detectors, 1-5 October 2006, Siena, Italy.}
\end{minipage}
\vspace{0.8mm}

The use of nuclear emulsions to record tracks of charged 
particles with an accuracy of better than 1 micron is possible in large physics experiments thanks to the recent improvements in the industrial production of emulsions and to the development of fast automated microscopes. 

The European Scanning System (ESS) is a fast automatic system
developed for the mass scanning of the emulsions of the OPERA 
experiment, which requires microscopes with scanning speeds of 
about 20 cm$^2$/h.
%
%In order to measure with high accuracy and high speed, very strict constraints 
%must be satisfied in terms of mechanical precisions, camera speed and image 
%processing power. 
%
Recent improvements in the technique and measurements with ESS 
are reported.
\vspace{-0.5mm}
\end{abstract}
\maketitle

%______________________________________________________________________________

\section{Introduction}
Nuclear emulsions were largely used during the last century and are connected to many discoveries in the early days of nuclear and particle physics. Since the production and the measurements were carried out manually, the amount of emulsions used for the experiments was quite small\,\cite{Powell}. 
 
Significant improvements in the emulsion technique and the development of fast automated scanning systems during the last two decades\,\cite{TS,SUTS,SYSAL} have made possible the use of nuclear emulsions in large scale experiments like
OPERA at the INFN Gran Sasso Underground Laboratories.

OPERA is a long baseline experiment\,\cite{OPERA1} designed to search for $\nu_{\mu} \rightarrow\nu_{\tau}$ oscillations in the parameter range suggested by atmospheric neutrino experiments\,\cite{Fukuda:1998mi,Ambrosio:1998wu,Allison:2005dt}. The goal is to observe the appearance of the short-lived $\tau$ leptons in a pure $\nu_\mu$ beam produced by the CNGS facility at CERN\,\cite{CNGS}. The $\tau$ lepton is directly identified through its decay topologies which need track position and angular measurements with accuracies of $\sim1$ micron and a few milliradiants respectively.

%Therefore high-accuracy tracking devices as nuclear emulsion are needed.

\section{Nuclear emulsions. OPERA Target}
Nuclear emulsions are made of micro-crystals of silver halides (AgBr) dispersed in a gelatin layer. The energy released by ionizing particles to the crystals, produces a latent image which is stable in time. A chemical development process reduces the irradiated grains to metallic Ag. After fixing and washing to remove undeveloped crystals the gelatin is transparent; the path of an ionizing particle is visible as a sequence of black silver grains about 0.5 $\mu$m in size \cite{Powell}.

In order to see the track of a particle in
emulsion, almost 30 developed grains every 100 $\mu$m of path are
necessary. The grains which accidentally
develop are randomly distributed in the emulsion volume and
their concentration should be $<5$ in 1000 $\mu$m$^{3}$.

In the OPERA detector, as in other recent experiments, the emulsions are used as thin films: a pair of emulsion layers (\ThickEmu $\mu$m thick) is mounted on both sides of a plastic base (\ThickBase $\mu$m thick) \cite{OpFilm}.

%\section{OPERA experiment}
   
\begin{figure}[ht]
\begin{center}
\includegraphics[width=17.5pc]{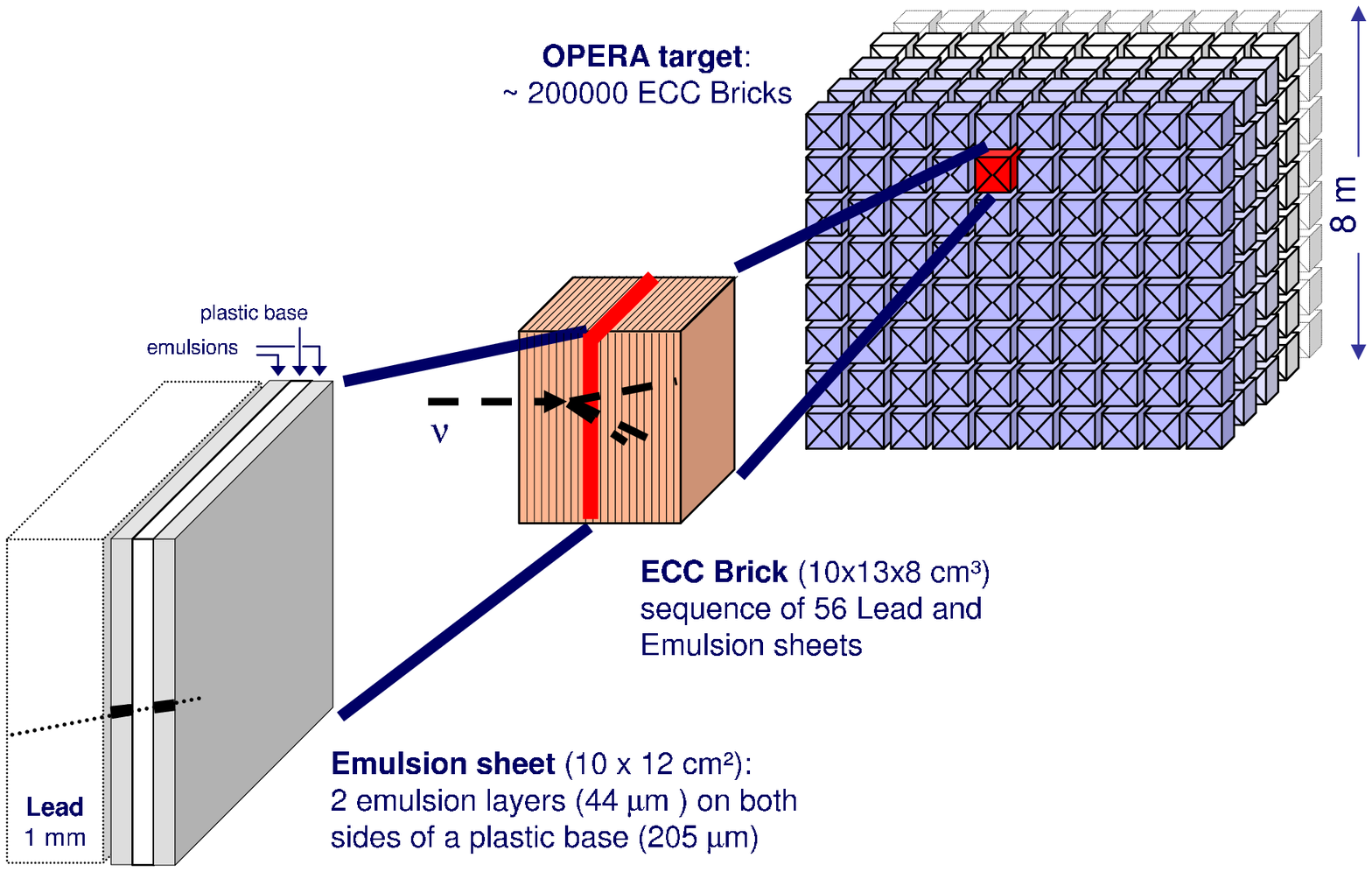}
\spacebeforefigurecaption
\caption{The OPERA target is segmented into $\sim200,000$ \textit{bricks} which are sequences of lead and emulsion sheets.\label{Fi:OperaTarget}}
\spaceafterfigurecaption
\end{center}
\end{figure}

%\subsection{OPERA Lead-Emulsion Target}
The OPERA detector is a hybrid system consisting of electronic detectors and a massive lead-emulsion target segmented into %$206,336$ 
$\sim200,000$ \textit{bricks} (Fig. \ref{Fi:OperaTarget}).

A \textit{brick} is a sequence of 56 lead sheets, acting as target, interleaved with 57 emulsion sheets, acting as high precision trackers,  and satisfies the need of both a large mass and a high precision tracking capability.\footnote{This technique is historically called \textit{Emulsion Cloud Chamber} (ECC).}
%Industrially mass production of 13 600 000 films by Fuji Co in 2 years.

With the CNGS neutrino beam\,\cite{CNGS} at its nominal intensity, $\sim 30$ neutrino selected interactions per day are expected. Therefore, $\sim 2000$ emulsion sheets per day must be (partially) scanned in order to find the vertex and analyze the event. In total, $\sim6000$\,$\mbox{cm}^2$ per day %($\sim200$\,$\mbox{cm}^2$ per brick) 
have to be analyzed with a sub-micrometric precision per 5 
years of data taking.
% ($\gtrsim 30000$ neutrino interactions). 

%Real time analysis: 
% $\sim$ 30 neutrino selected interactions per day;
% $\sim$ 6000 cm$^2$ per day have to be analyzed with a sub-micrometric precision during 5 years data taking.
 
%\includegraphics[width=15pc]{opera_target}
 
\section{Automatic Emulsion Scanning System}
New automatic fast automatic scanning systems have been developed: the European Scanning System (ESS) and the S-UTS in Japan.

\begin{figure}[ht]
\begin{center}
\includegraphics[width=17.5pc]{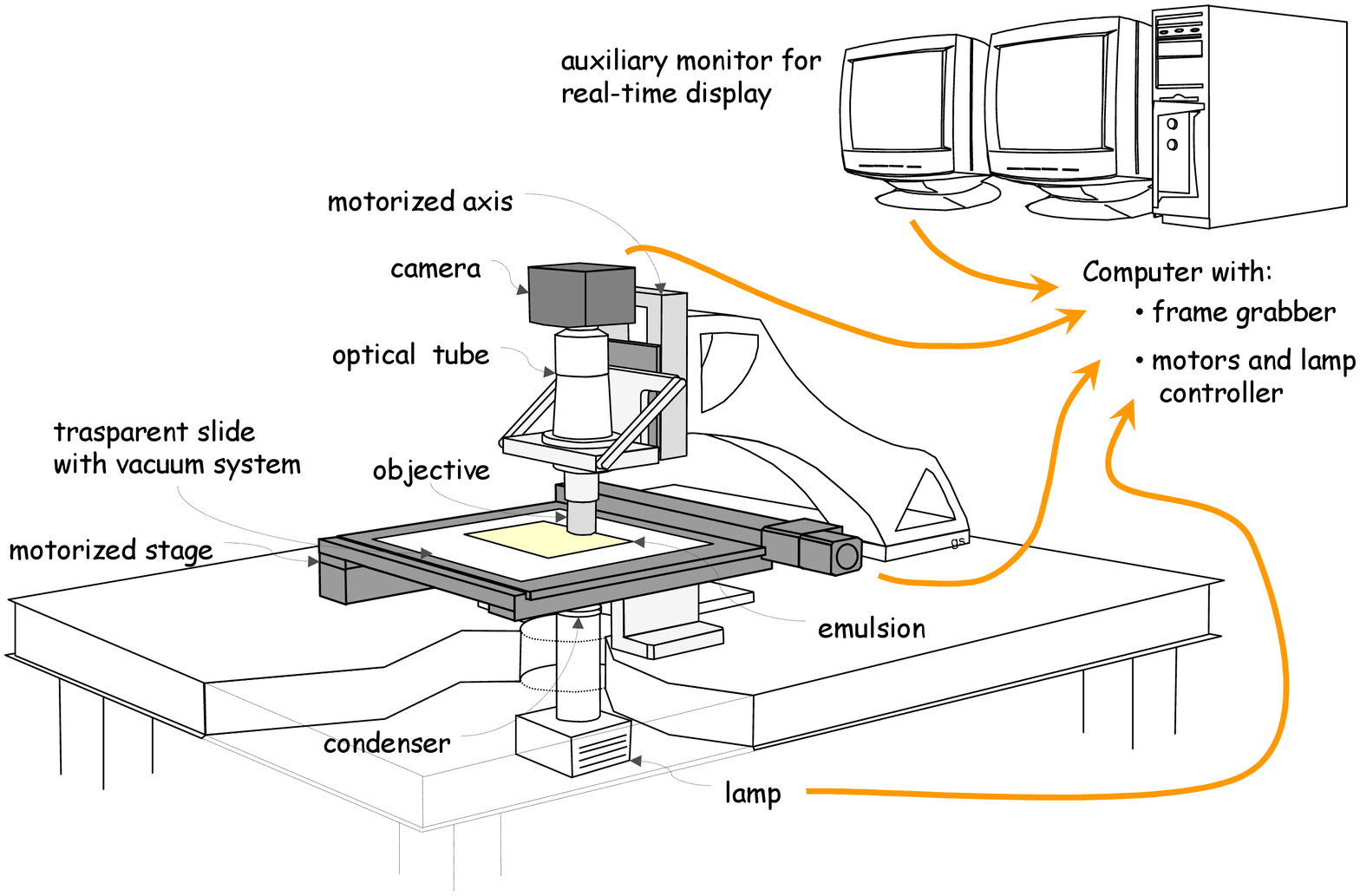}
\spacebeforefigurecaption    
\caption{Schematic layout of the European Scanning System microscope. \label{Fi:ESSLayout} }
\spaceafterfigurecaption
\end{center} 
\end{figure}

\begin{figure}[ht]
\begin{center}
\includegraphics[width=16.5pc]{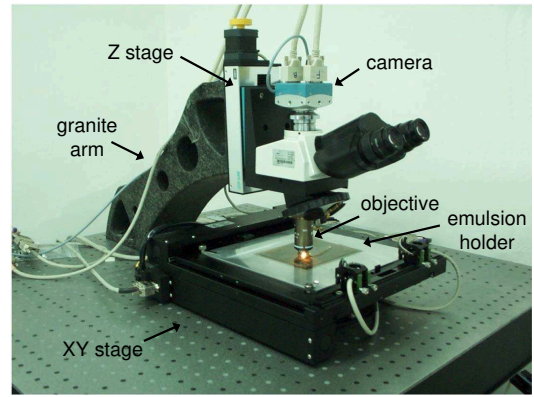}
\spacebeforefigurecaption  
\caption{A photograph of one of the microscopes of the European Scanning System.\label{Fi:ESS}}
\spaceafterfigurecaption
\end{center}
\end{figure}

The ESS (Fig. \ref{Fi:ESSLayout} and Fig. \ref{Fi:ESS}) is based on the use of commercial hardware components or developed in collaboration with specialized companies. 
The ESS is able to scan an emulsion volume of 44 $\mu$m thickness with a speed of 20 cm$^2$/h. This represents an improvement of more than an order of magnitude with respect to the systems developed in the past. 
The Japanese S-UTS system uses a dedicated hardware suitable for point scanning with a speed of 1.2 s/prediction ($\sim$ 15 min/brick).

%\section{Emulsion Readout}
By adjusting the focal plane of the objective, the whole 44 $\mu$m emulsion thickness is spanned and a sequence of 15 tomographic images of each field of view, taken at equally spaced depth levels (3 $\mu$m), is obtained. 
Emulsion images are digitized, converted into a grey scale of 256 levels, sent to a vision processor board, hosted in the control workstation, and analyzed to recognize sequences of aligned grains (clusters of dark pixels of given shape and size).

\begin{figure}[ht]
\begin{center}
\includegraphics[width=17.5pc]{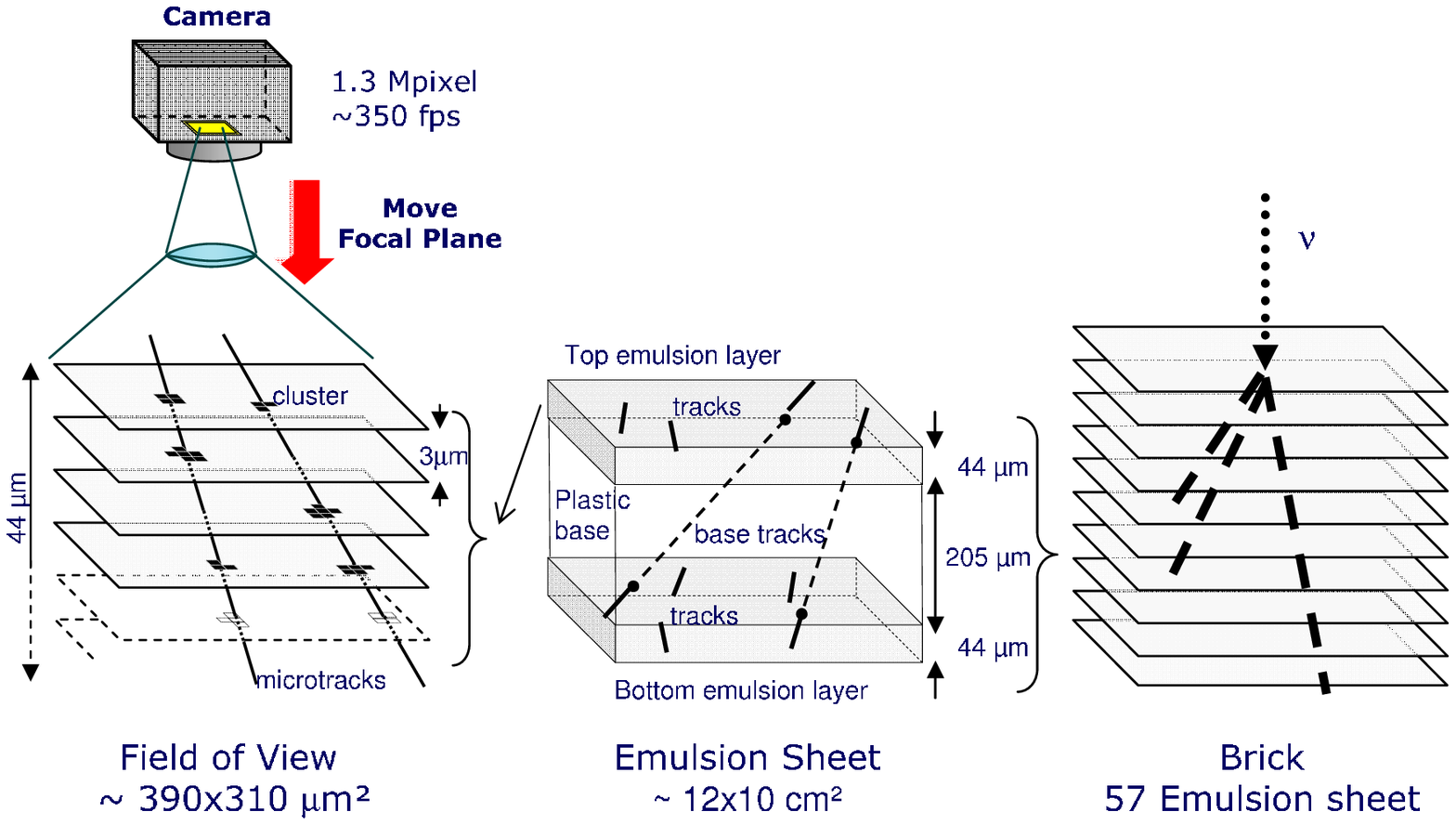}
\spacebeforefigurecaption
\caption{Track reconstruction: for each field of view, several emulsion images are taken by moving the optical axis and track segments are found by connecting aligned grains (\textit{left}). Tracks are reconstructed by linking both sides of the emulsion sheet (\textit{center}) and then all sheets of the entire brick (\textit{right}).\label{Fi:TrackReco}}
\spaceafterfigurecaption
\end{center}
\end{figure}

The three-dimensional structure of a track in an emulsion layer is reconstructed by combining clusters belonging to images at different levels and searching for geometrical alignments. A linear fit to these clusters allows the determination of the track position and angle. 
After emulsion sheets alignment, tracks are reconstructed in the entire brick (Fig. \ref{Fi:TrackReco}).

\section{Scanning performances}
\begin{figure}[ht]
\begin{center}
\includegraphics[clip,width=36.5mm]{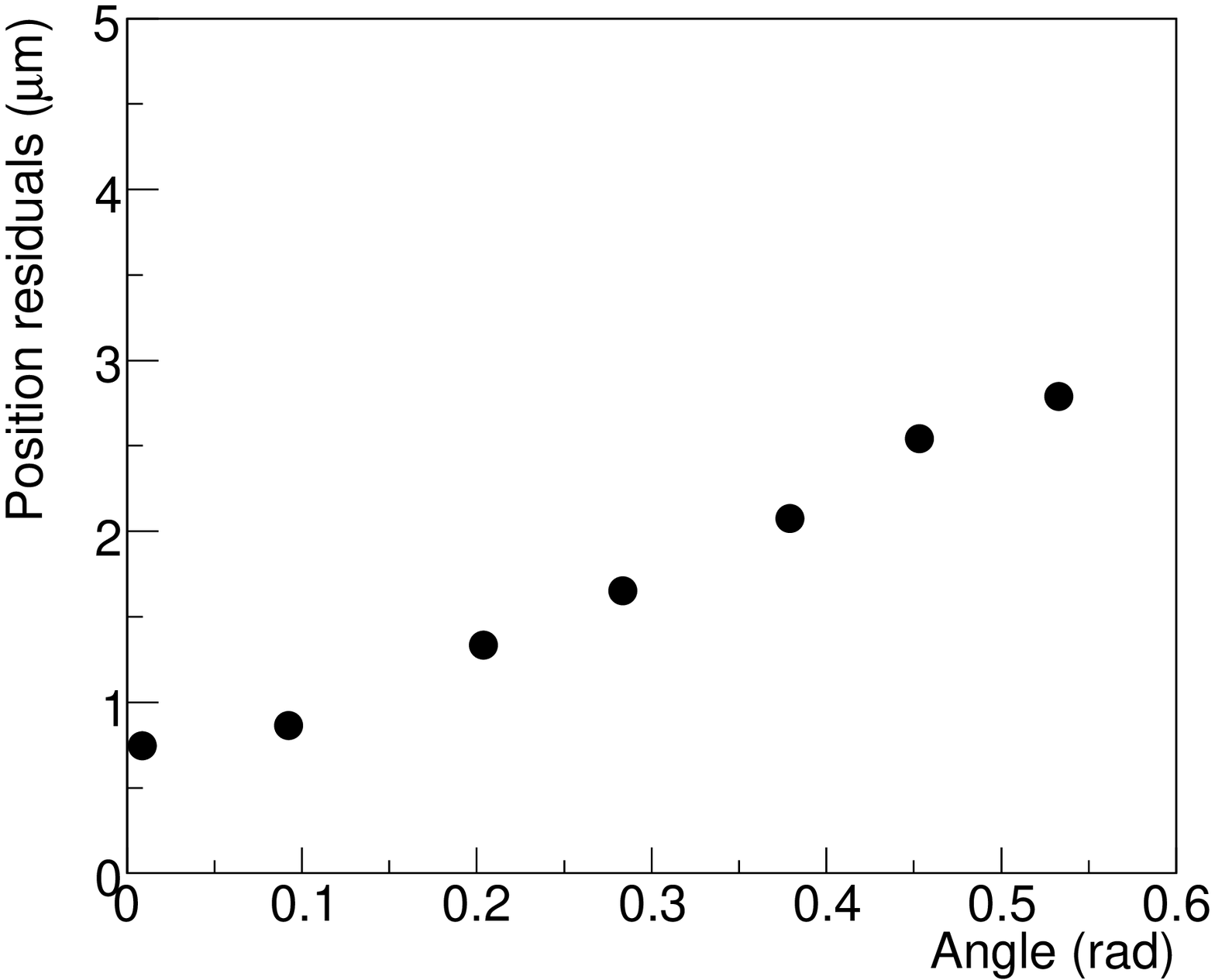}\hspace{-0.5mm}
\includegraphics[clip,width=36.5mm]{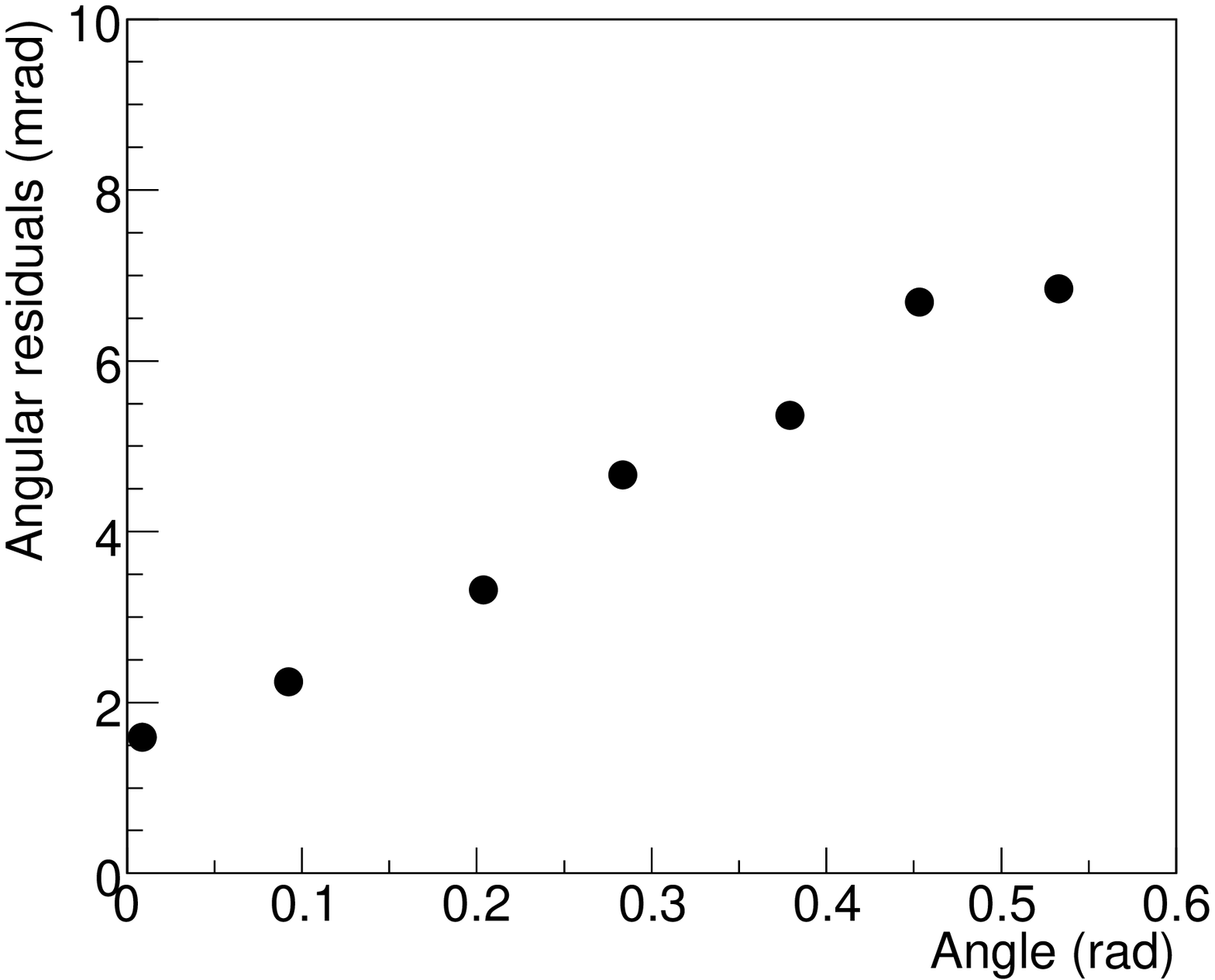}
\spacebeforefigurecaption  
\caption{The position resolution of base tracks as function of the reconstructed angle (\textit{left}). The angle resolution of base tracks (\textit{right}).  
%It is evaluated by comparing base track angles with respect to the volume track angles. 
The errors (that are inside the dimensions of each black point) are only statistical.\label{Fi:VolumeResAng}}
\spaceafterfigurecaption
\end{center}
\end{figure}

Several test exposures at pions beams were performed to estimate the scanning performances. 
The scanning systems are successfully running with high efficiency ($>$90\%), good signal/background ratio ($\sim 2\;\mathrm{tracks/cm}^2 /[\mathrm{angle}<0.4\;\mathrm{rad}]$) and the design speed of 20 cm$^2$/h.  Position and angular resolutions at small incident angles are $\sigma_{position}$ = 1 $\mu$m and $\sigma_{angle}$ = 2 mrad (Fig. \ref{Fi:VolumeResAng}).

\section{Conclusions}
The features and performances of the European Scanning System (ESS) have been described. The resulting tracking efficiencies have been evaluated to be above 90\% in the $[0, 600]$ mrad angular range with resolutions of $\sim 1$ $\mu$m and $\sim 1$ mrad for vertical tracks.

The ESS has reached the speed of $\sim20$ cm$^2$/h in an emulsion volume \ThickEmu\ $\mu$m thick. This represents an improvement of more than an order of magnitude with respect to the systems developed in the past. The scanning performances satisfy the requirements of the OPERA experiment.

About 20 ESSs have been installed in European laboratories collaborating in the OPERA experiment. Five more have been installed at the Gran Sasso Laboratory (LNGS).

\end{document}